\documentclass[]{spie}  

\usepackage{amsmath,amsfonts,amssymb}
\usepackage{graphicx}

\title{Data processing on simulated data for SHARK-NIR}

\author[a,b]{Carolo E.}
\author[a,b,c]{Vassallo D.}
\author[a,b]{Farinato J.}
\author[e,b]{Agapito G.}
\author[a,b]{Bergomi M.}
\author[d]{Carlotti A.}
\author[a,b]{De Pascale M.}
\author[a,b]{D'Orazi V.}
\author[a,b]{Greggio D.}
\author[a,b]{Magrin D.}
\author[a,b]{Marafatto L.}
\author[a,b]{Mesa D.}
\author[e,b]{Pinna E.}
\author[e,b]{Puglisi A.}
\author[f,b]{Stangalini M.}
\author[d]{Verinaud C.}
\author[a,b]{Viotto V.}
\author[a,b]{Biondi F.}
\author[a,b]{Chinellato S.}
\author[a,b]{Dima M.} 
\author[e,b]{Esposito S.}
\author[f,b]{Pedichini F.}
\author[a,b]{Portaluri E.}
\author[a,b]{Ragazzoni R.}
\author[a,b,c]{Umbriaco G.}

\affil[a]{INAF - Osservatorio Astronomico di Padova, Vicolo dell'Osservatorio 5, 35122, Padova, Italy}
\affil[b]{ADONI - Laboratorio Nazionale Ottiche Adattive, National Laboratory for Adaptive Optics, Italy}
\affil[c]{Dipartimento di Fisica e Astronomia, Universit\`a degli Studi di Padova, Vicolo dell'Osservatorio 3,
35122, Padova, Italy}
\affil[d]{Institut de Plan\'etologie et d'Astrophysique de Grenoble, 414, Rue de la Piscine, Domaine Universitaire, 38400 St-Martin d'H\`eres, France}
\affil[e]{INAF - Osservatorio Astrofisico di Arcetri, Largo Enrico Fermi 5, 50125 Firenze, Italy}
\affil[f]{INAF - Osservatorio Astronomico di Roma, Via Frascati 33, 00078 Monte Porzio Catone, Roma, Italy}

\authorinfo{Further author information: (Send correspondence to E.C.)\\E.C.: E-mail: elena.carolo@oapd.inaf.it}

\begin{document} 
\maketitle

\begin{abstract}
A robust post processing technique is mandatory to analyse the coronagraphic high contrast imaging
data. Angular Differential Imaging (ADI) and Principal Component Analysis (PCA) are the most
used approaches to suppress the quasi-static structure in the Point Spread Function (PSF) in order to
revealing planets at different separations from the host star.
The focus of this work is to apply these two data reduction techniques to obtain the best limit
detection for each coronagraphic setting that has been simulated for the SHARK-NIR, a
coronagraphic camera that will be implemented at the Large Binocular Telescope (LBT).
We investigated different seeing conditions ($0.4"-1"$) for stellar magnitude ranging from R=6 to
R=14, with particular care in finding the best compromise between quasi-static speckle subtraction
and planet detection.
\end{abstract}


\section{INTRODUCTION}
\label{sec:intro}  
SHARK is a coronagraphic camera proposed for Large Binocular Telescope in the framework of the “2014 Call for Proposals
for Instrument Upgrades and New Instruments”~\cite{2014ebi..confP4.74F, 2015IJAsB..14..365F, 2016SPIE.9909E..31F, 2016SPIE.9911E..27V}.
We are building this tool because first of all we hold an excellent adaptive optic (AO) performance, we are in the northern hemisphere with a strong scientific case and the purpose is going on sky in a very short time. 
In order to take advantage of these points we proposed a simple camera which will allow direct imaging, coronagraphic imaging and coronagraphic low resolution spectroscopy. 
SHARK-NIR together with the SHARK-VIS channel, are covering a wide wavelength domain, going from 0.6$\mu m$ to 1.7$\mu m$ (Y to H band). 
SHARK-NIR will offer extreme AO direct imaging capability on a field of view (FoV) of about 18\textquotedblright x 18\textquotedblright, and a simple coronagraphic spectroscopic mode offering spectral resolution ranging from 100 to 700. \\
The main science case of SHARK-NIR is searching for giant planets, to succeed, an high contrast is necessary.  
We also emphasize that the LBT AO SOUL upgrade will further improve the AO performance, making possible to extend the exoplanet search to target fainter than normally achieved by other 8-m class telescopes, and opening in this way to other very interesting scientific scenarios, such as the characterization of AGN and Quasars, normally too faint to be observed, and increasing considerably the sample of disks and jets to be studied. 

\section{The Angular Differential Imaging: ADI}
\label{sec:adi} 
The Angular Differential Imaging~\cite{adi} is a post processing technique used on direct imaging data to suppress the quasi-static structure presents in the PSF~\cite{Marois}. The acquisition of a set of images, usually up to a few hundreds, is performed with the instrument rotator turned off. In this way, the quasi-static PSF is stable during the observation while a rotation of the FoV with respect the instrument occurs. 
The algorithm is structured in three simple steps: 
first, a reference PSF is generated by median combining all images in the sequence and it is subtracted from each image to remove the quasi-static structure. 
Then image differences are de-rotated to align the FoV. 
Finally, de-rotated images are median combined. \\
The peculiarity of the data reduction pipeline we written is the implementation of different variants of the reference PSF. We taken into account a single-median as reference PSF as in the classical ADI of course, but also multiple-medians. In this latest case the images sequence is divided into two or more subsets
and the single-median subtraction is applied separately for each
of them. In this way, reference PSFs are closer in time (and hence
more correlated) to the images they are subtracted from. The
drawback is the enhanced planet cancellation (see Sect.~\ref{sec:selfsub}). In addition we also performed the Principal Component Analysis (PCA), this algorithm is based on a
statistical representation of each frame as a linear combination of
its principal orthogonal components. These components are estimated
by diagonalization of the covariance matrix associated to
the signal. Its application has been reported to be very effective
particularly by manipulation of the number of principal components
to maximize the signal from the planet near its host star~\cite{pca, pca2}. 
The number of modes used for the PCA analysis is associated
to the variance of the corresponding principal component.

\section{Self-subtraction of the planet light}
\label{sec:selfsub}
\begin{figure} [ht]
   \begin{center}
   \begin{tabular}{c} 
   \includegraphics[height=8cm]{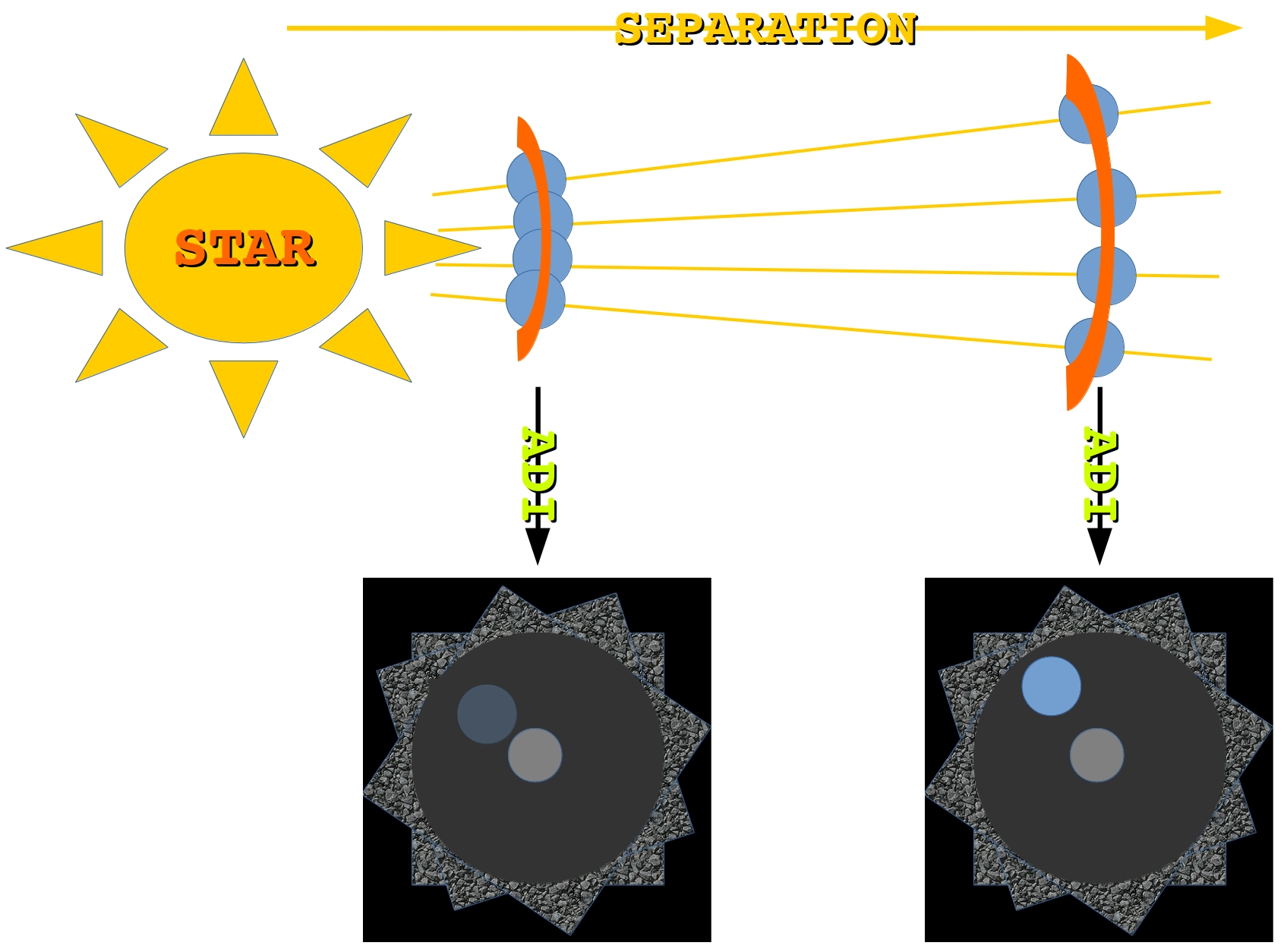}
   \end{tabular}
   \end{center}
   \caption[example] 
   { \label{fig:selfsubscheme} 
Graphic scheme of the self-subtraction effect depends on the separation from the host star.}
   \end{figure}

Using small subsets to generate the reference PSF can help attenuating the speckle noise, but it also results in a growing risk of planet removal if not enough field rotation occurs in the subset itself. 
This stresses out how important is to account for the planet self-subtraction~\cite{carolo}. 
In general, self-subtraction depends on the planet separation from the host star: the drop of its light is bigger at small separations and decreases at large separations (see Fig.~\ref{fig:selfsubscheme}). 
This is due to the fact that the same FoV rotation corresponds to a slower motion of the planet at small radial separations from the star, with respect to larger separations. 

\subsection{The effect of the self-subtraction}
\label{sec:selfsubeffect}
\begin{figure} [ht]
   \begin{center}
   \begin{tabular}{c} 
   \includegraphics[height=9cm]{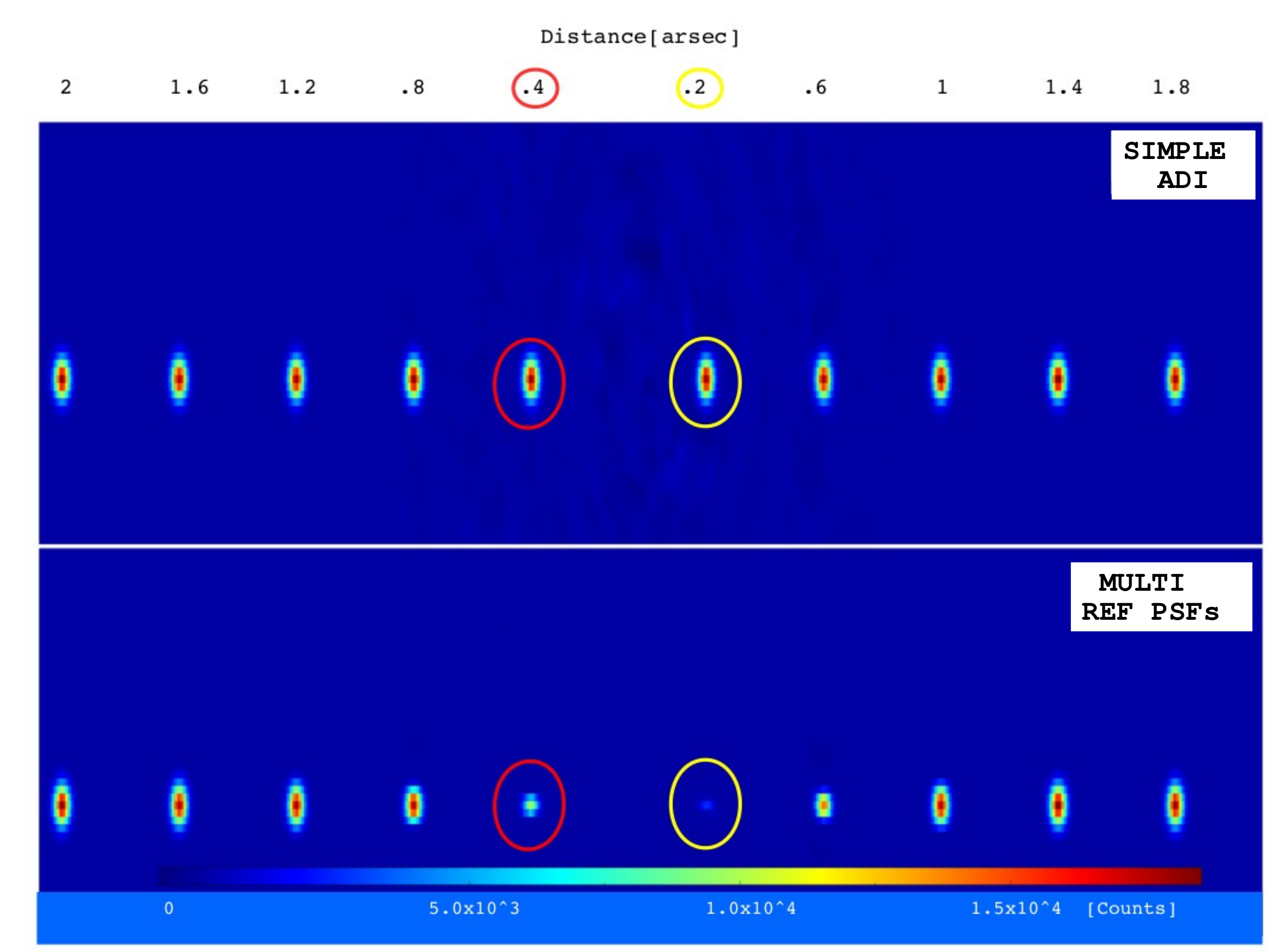}
   \end{tabular}
   \end{center}
   \caption[example] 
   { \label{fig:effect} 
The effect of the self-subtraction on the planet light, it is stronger near the host star. A simple ADI case (upper panel) and a more reference PSFs subtraction (lower panel) in comparison.}
   \end{figure}

This effect of the self-subtraction results in a subtraction of both speckles and the planet signal near to the star. In order to account for this effect, the pipeline injects fake planets in simulated images at different separations and position angles. The simulated datacube in which we injected the fake planets is processed by the pipeline and the post-processed result is analysed to obtain the quantity of the planet light lost (see Fig.~\ref{fig:effect}).

\subsection{The field of view rotation at LBT}
\label{sec:fovrot}
\begin{figure} [ht]
   \begin{center}
   \begin{tabular}{c} 
   \includegraphics[height=10cm]{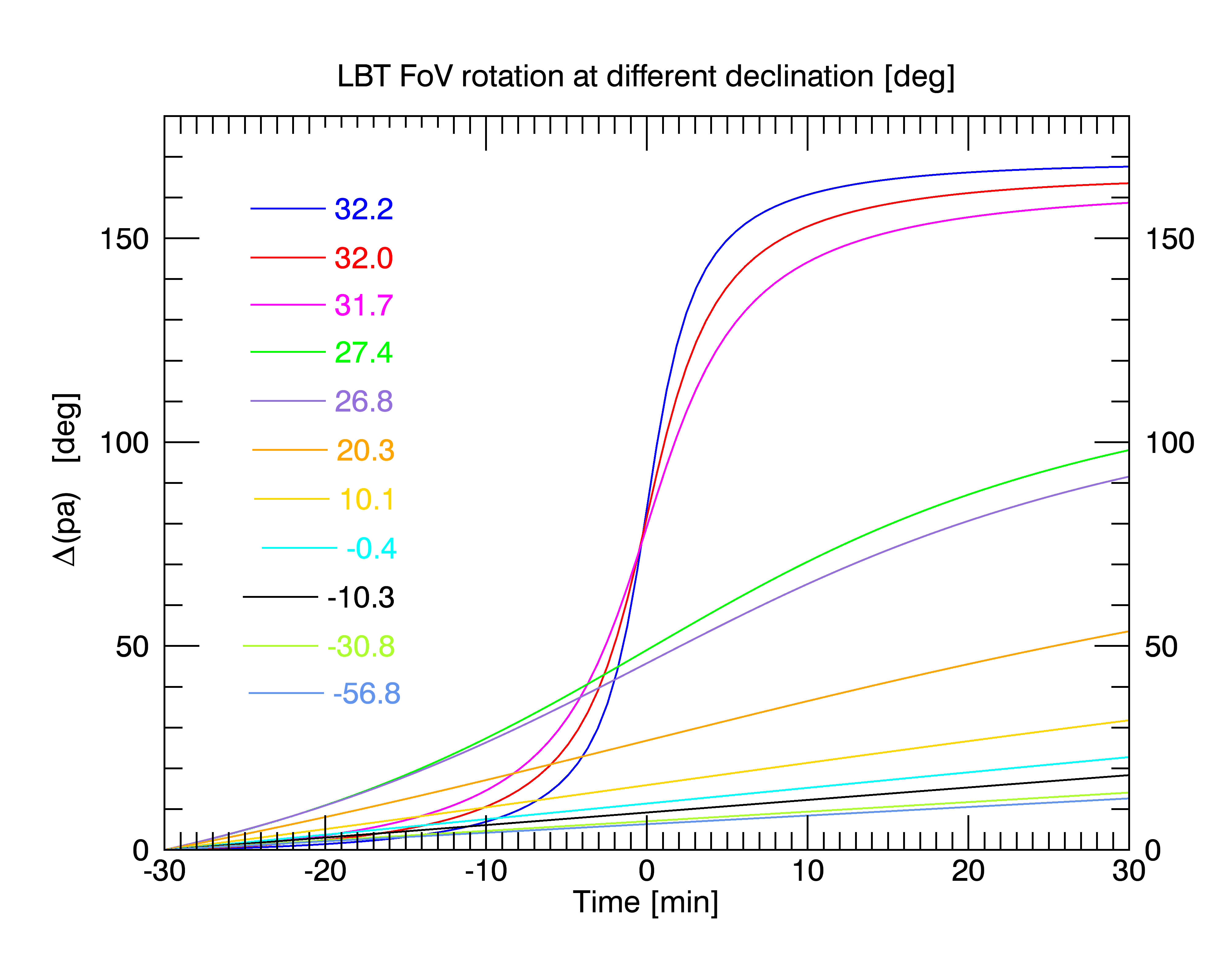}
   \end{tabular}
   \end{center}
   \caption[example] 
   { \label{fig:fovrot} 
An example of parallactic angle values for objects at different declinations during an observation of 1 hour at
LBT.}
   \end{figure}

To study this effect we used a sequence of 30 images, assuming they cover uniformly a space of time of 1 hour. Two FoV rotation cases (30$^\circ$ and 90$^\circ$) are explored, according to the LBT object visibility (see Fig.~\ref{fig:fovrot}). We also assumed that star culmination occurs in the middle of the observation. 
For the simulation we chose 90 degree of FoV rotation corresponding to an object declination of 27$^\circ$, more or less the Taurus-Auriga star forming region.

\subsection{Estimation of the signal loss}
\label{sec:selfsubquantify}
\begin{figure} [ht]
   \begin{center}
   \begin{tabular}{c} 
   \includegraphics[height=8cm]{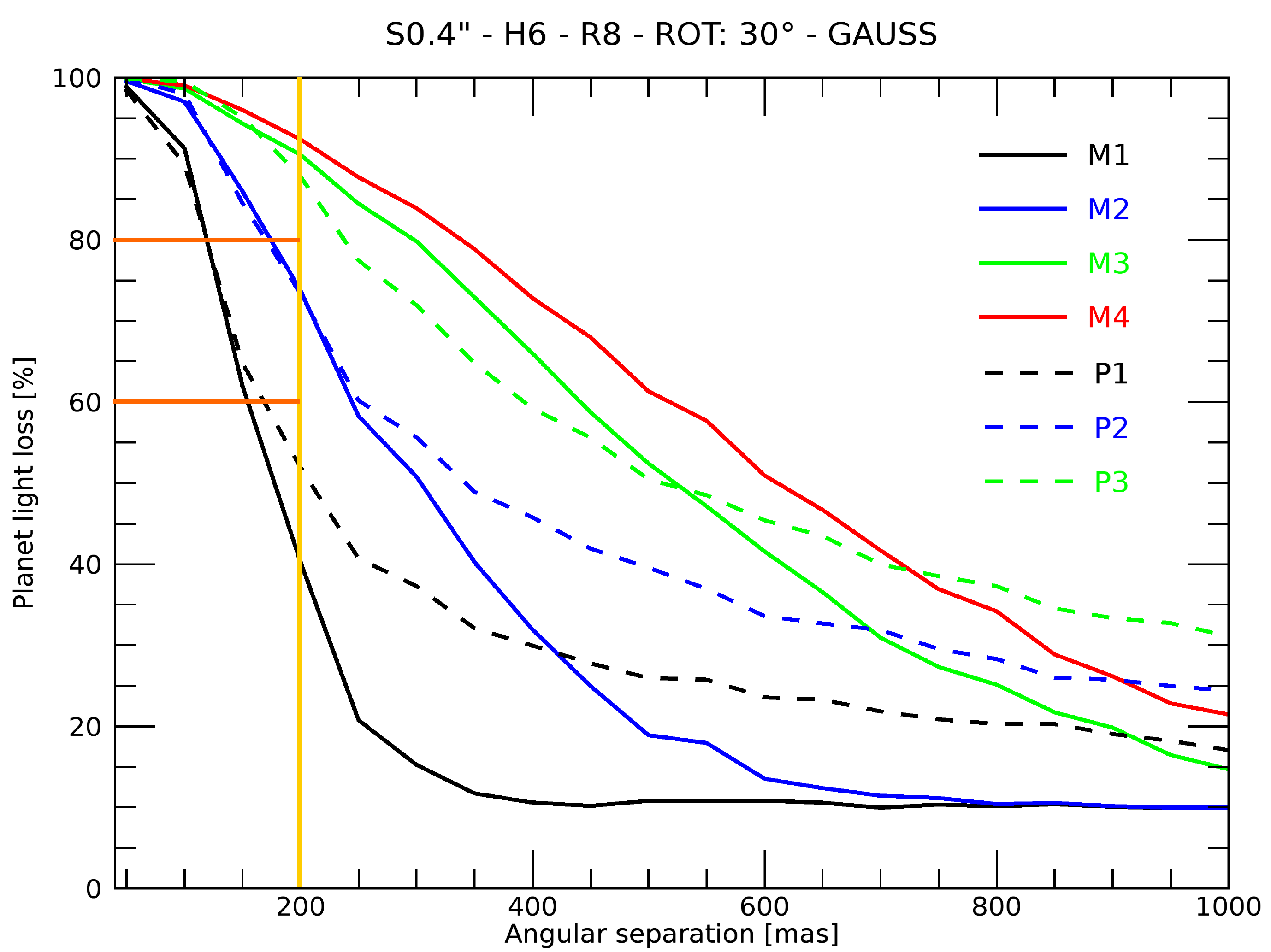}
   \end{tabular}
   \end{center}
   \caption[example] 
   { \label{fig:ss30} 
The planet light subtracted for a 30$^\circ$ FoV rotation. Amount of the planet light that is self-subtracted by the post processing reduction (M1 corresponds to simple ADI, M2 is an ADI by using two references psf and so on. For the PCA the capital P is associated to the number of used modes) for Gaussian-Lyot coronagraph in a high Strehl condition.}
   \end{figure}

\begin{figure} [ht]
   \begin{center}
   \begin{tabular}{c} 
   \includegraphics[height=8cm]{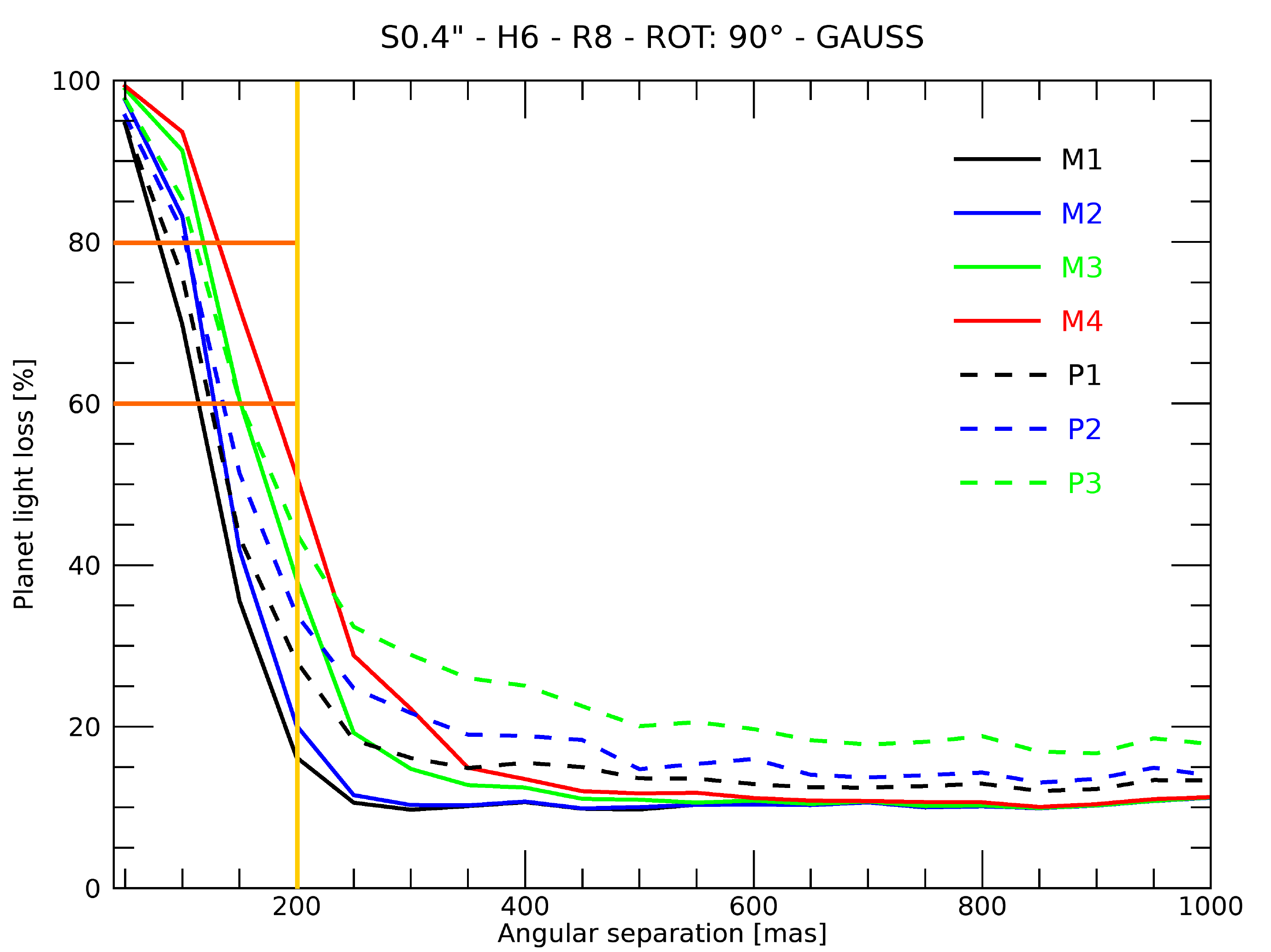}
   \end{tabular}
   \end{center}
   \caption[example] 
   { \label{fig:ss90} 
The planet light subtracted for a 90$^\circ$ FoV rotation. Amount of the planet light that is self-subtracted by the post processing reduction (M1 corresponds to simple ADI, M2 is an ADI by using two references psf and so on. For the PCA the capital P is associated to the number of used modes) for Gaussian-Lyot coronagraph in a high Strehl condition.}
   \end{figure}

To quantify this effect we compared the residual (post-ADI) signal to the initial one, in the way to generate a cancellation profile. 
In the Figure \ref{fig:ss30} and Figure \ref{fig:ss90} we report some examples of these profiles: signal loss is plotted as a function of angular separation from the star for different ADI variants. 
For a FoV rotation of 30$^\circ$, single-median subtraction and PCA using 1 mode preserve almost 60$\%$ of the signal for separations larger than 200mas (see Fig.~\ref{fig:ss30}). 
As FoV rotation increases (90$^\circ$ as in the figure on the right), a growing number of modes for PCA subtraction can be used (see Fig.~\ref{fig:ss90}). 
At very small angular separations (150-200mas), only single-median subtraction and 1-mode PCA in the case of 90$^\circ$ FoV rotation allow to preserve almost the 60$\%$ of the signal. 

\section{Post-processing code}
\label{sec:ppcode}
Simulations~\cite{Vassallo} are based on the Fresnel end-to-end propagation. 
We used closed-loop atmospheric residuals for different seeing conditions ($0.4"-1"$) and guide star magnitudes (from R=6 to
R=14). We take into account Non Common Path Aberrations and Telescope Vibrations and polychromatic imaging was also developed. 
The result of each coronagraphic simulation  is a datacube of 30 frames.
The pipeline used to analyse these simulated data takes into account the cancellation factor by injecting fake planets in the data at different position.\\
For the detection limit, the metric we adopt is the 5-$\sigma$ detection limit: 
we compute the standard deviation of counts in the final ADI-processed frame as a function of angular separation. 
The standard deviation is calculated pixel-to-pixel: for each individual pixel in the final ADI-processed frame, we select two close-by regions for computation. 
This is the the same procedure used in simulating planet detection for SPHERE~\cite{Sphere, spheretec}. 
The output of the procedure is a bi-dimensional noise map, from which we extract a radial profile via azimuthal averaging. 
Standard deviation is then multiplied by five and divided to the peak of the field-nobstructed PSF, for sure taking into account the self-subtraction factor.
The self-subtraction value is calculated as:
\begin{equation}
SS = (P_{ADI} - Back)/P_i 
\end{equation}
Where $P_{ADI}$ is the intensity of the injected fake planet after the post processing code was applied, $Back$ is the background value calculated on a frame edge and $P_i$ is the initial intensity of the fake planet we injected. These values are evaluated for each planet at different separation in a way to generate a cancellation profile depending on the distance from the host star.

\section{Performance in high Strehl condition at different jitter values}
\label{sec:highstrehl}
Vibrations arise from resonant modes in the structure of telescope, in particular the swing arm supporting the Adaptive Secondary Mirror. 
These modes are excited by wind shacking and/or telescope tracking and mainly introduce tip and tilt aberrations which are only partially filtered out from the AO system. 
We indicate jitter as the amount in rms of the residuals after AO-correction (J3 = 3mas rms and J10 = 10mas rms).

\subsection{Coronagraphic configurations}
\label{sec:coro1}
\begin{figure} [ht]
   \begin{center}
   \begin{tabular}{c} 
   \includegraphics[height=10cm]{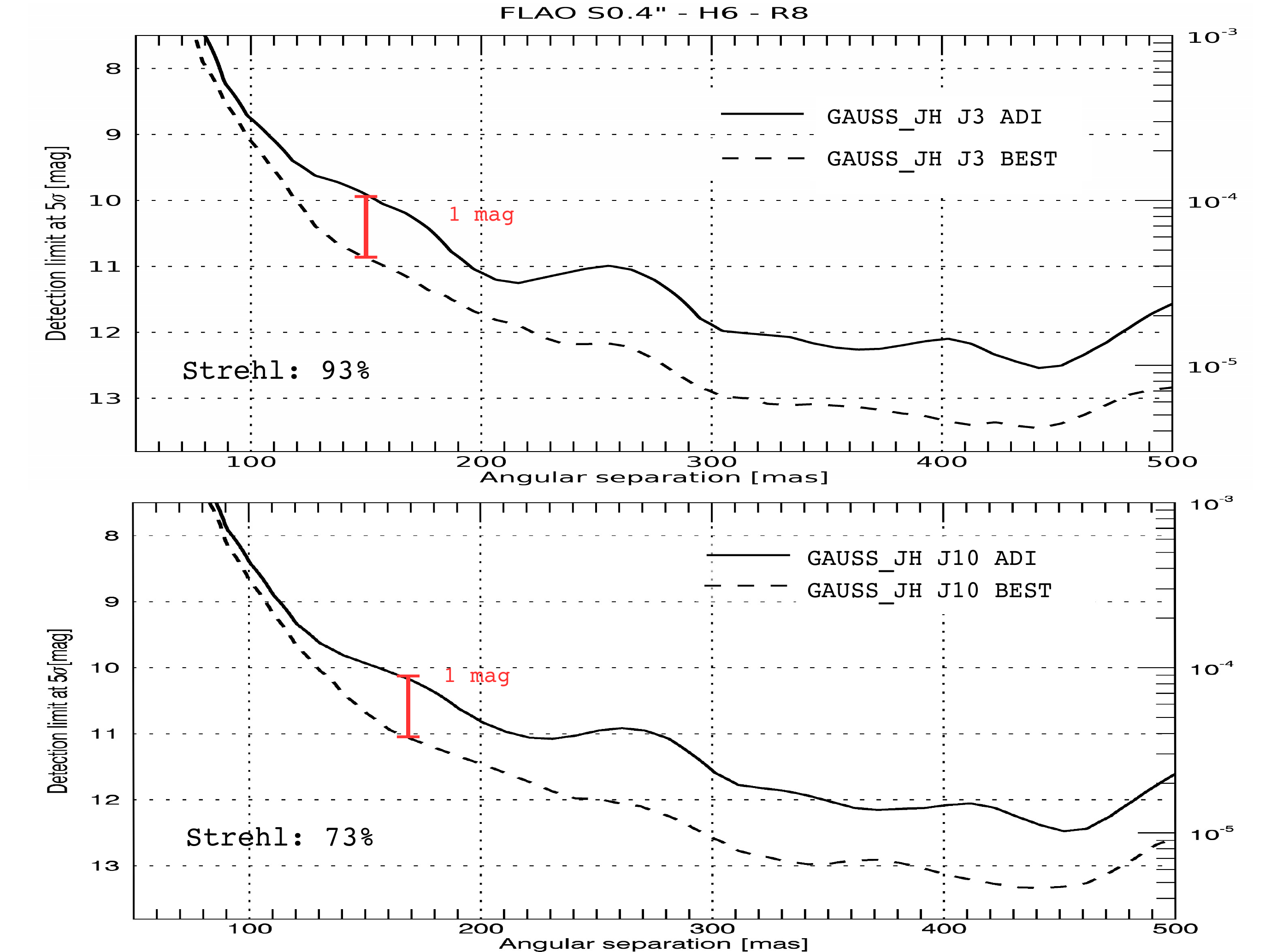}
   \end{tabular}
   \end{center}
   \caption[example] 
   { \label{fig:gaussh} 
Detection limit for a R=8 magnitude star and seeing $0.4"$ for the Gaussian-Lyot coronagraph. In both panel the performance of a simple ADI (solid line) and of the pipeline in comparison (dashed line). In the upper panel a low jitter value case, in the lower one a jitter value of 10mas rms. The solid red line represents the gain of the post processing code.}
   \end{figure}

\begin{figure} [ht]
   \begin{center}
   \begin{tabular}{c} 
   \includegraphics[height=10cm]{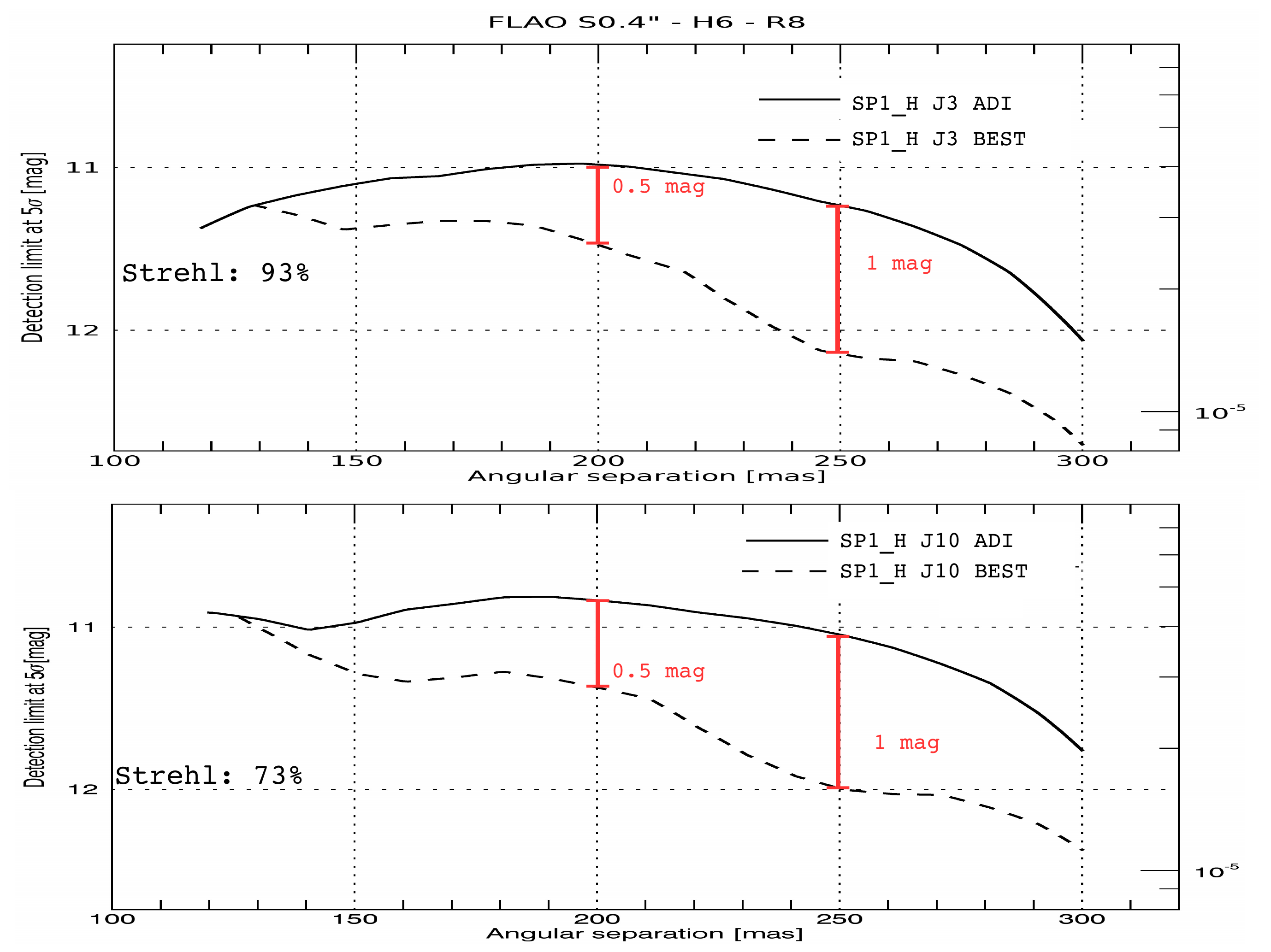}
   \end{tabular}
   \end{center}
   \caption[example] 
   { \label{fig:sph} 
Detection limit for a R=8 magnitude star and seeing $0.4"$ for a Shaped Pupil mask. In both panel the performance of a simple ADI (solid line) and of the pipeline in comparison (dashed line). In the upper panel a low jitter value case, in the lower one a jitter value of 10mas rms. The solid red line represents the gain of the post processing code.}
   \end{figure}

\begin{figure} [ht]
   \begin{center}
   \begin{tabular}{c} 
   \includegraphics[height=10cm]{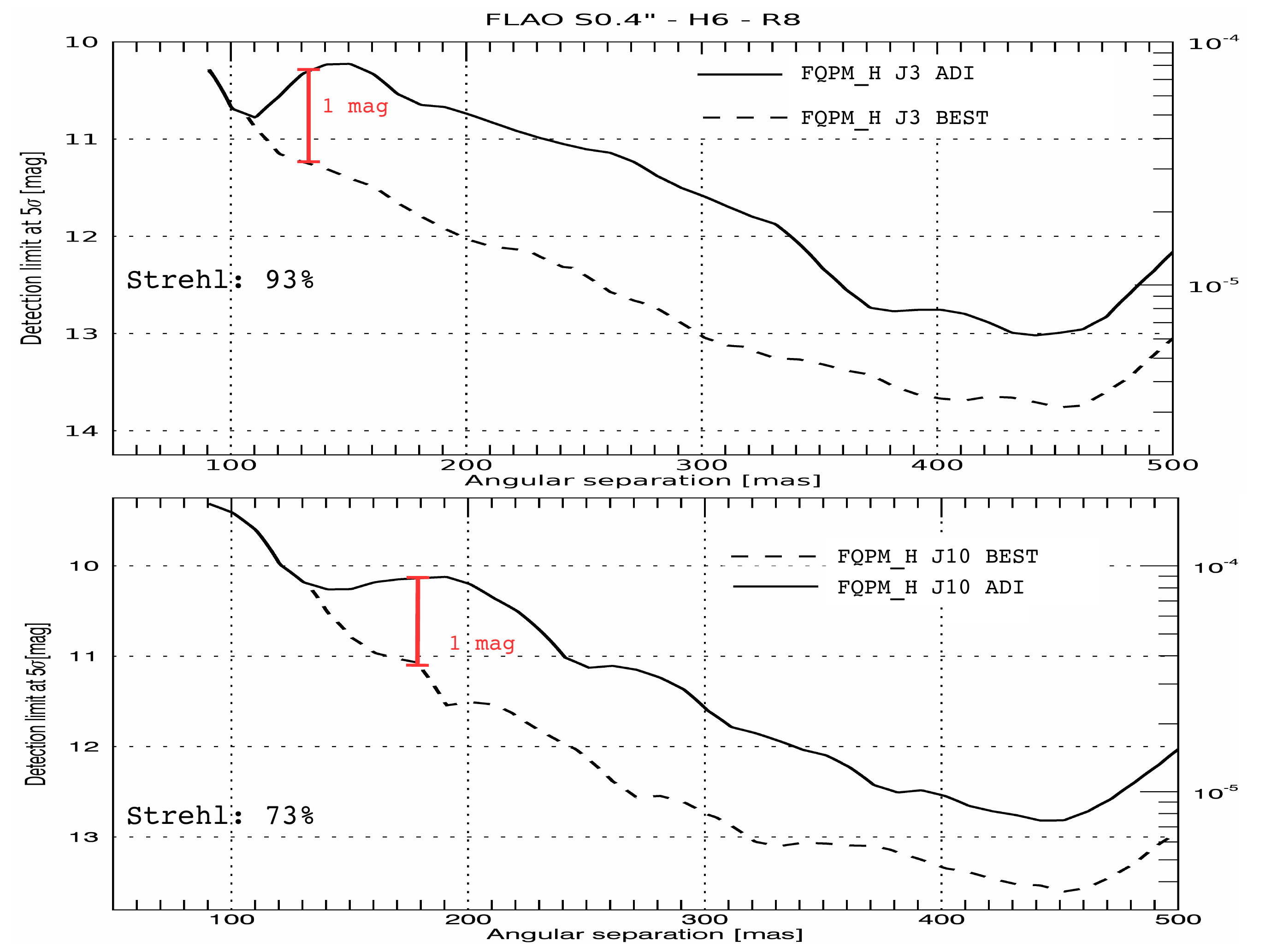}
   \end{tabular}
   \end{center}
   \caption[example] 
   { \label{fig:fqpmh} 
Detection limit for a R=8 magnitude star and seeing $0.4"$ for the Four Quadrant Phase mask. In both panel the performance of a simple ADI (solid line) and of the pipeline in comparison (dashed line). In the upper panel a low jitter value case, in the lower one a jitter value of 10mas rms. The solid red line represents the gain of the post processing code.}
   \end{figure}

We tested several coronagraphic masks. 
First of all the Gaussian-Lyot Coronagraph; it is similar to the Lyot coronagraphs~\cite{lyot}: instead of a hard edge mask, it modulates the amplitude of the electric field in the FP with a Gaussian filter. 
A Lyot stop is then put in the subsequent pupil plane. This solution allows to achieve a better contrast at small angular separations with respect to the classical Lyot configuration. In the Figure \ref{fig:gaussh} the coronagraph is indicated as $GAUSS JH$. \\
We also studied the performance of the Shaped Pupil Masks~\cite{sp}; classical SP simply consists of a binary transmission pattern applied in the pupil plane to generate a high contrast region in the subsequent focal plane. If the plane in which the high contrast zone is generated is not the detector plane, as in SHARK-NIR case, then a hard edge mask can be placed to remove all light falling out of this zone. Finally, a stop is placed in the subsequent pupil plane to remove residual light diffracted at the border of the pupil by the FP mask, as in the classical Lyot configuration. 
The coronagraph can generate high contrast all around the star (360$^\circ$ extent) or in two regions of variable extent (asymmetric discovery space). In the Figure \ref{fig:sph} the coronagraph is indicated as $SP1 H$. \\
We are considering the implementation of a fast tip-tilt correction during the exposure: this system allows us to evaluate focal plane coronagraphs requiring high PSF stability, such as the Four Quadrant Phase Mask (FQPM)~\cite{fqpm}. 
This coronagraph suppresses on-axis starlight by means of a phase mask inducing a phase shift on specific areas of the focal plane. In particular, the mask is arranged according to a four quadrant pattern: two quadrants on one diagonal without phase shift and the two other quadrants providing a π phase shift. 
Provided that the image of the star is exactly located at the center of the mask, the four beams combine in a destructive way and the stellar light is mostly rejected outside of the pupil area. Then a Lyot stop is placed in this exit pupil to remove the diffracted starlight. In the Figure \ref{fig:fqpmh} the coronagraph is indicated as $FQPM H$. \\
In the Figure \ref{fig:gaussh}, Figure \ref{fig:sph} and Figure \ref{fig:fqpmh} the results of the data reduction pipeline are presented. We take into account the self-subtraction factor and compare a simple ADI performance and the best post processing, by using median subtraction and PCA, at each separation from the host star. \\
We stress the gain by choosing for each distance the best post processing and preserving almost the 60$\%$ of the planet light simultaneously. In the Gaussian-Lyot and FQPM configuration the code performs a 1 magnitude better then a simple ADI near the Inner Working Angle (IWA), at separation lower then 200mas from the host star, both in condition of high and lower jitter values. While for the SP this gain is an half magnitude at 200mas and of 1 magnitude at 250mas in both jitter cases.

\section{CONCLUSION}
\label{sec:conclusion}
After having tested other kind of coronagraphs, as the asymmetric SP, the Apodized Phase Plate and the Vortex, we moved for choosing the final configuration of the SHARK-NIR. 
The asymmetric masks are attractive because of the good performance near the IWA. For these reasons, the final choice is to implement a symmetric SP and a couple of asymmetric masks. 
Since asymmetric masks cannot access the entire 360$^\circ$ around the star, they may be used for characterization of known objects. However, they are designed to generate two symmetric 110° high-contrast regions (for a total of 220$^\circ$ discovery space each) which are perpendicular to each other. Hence, if used in sequence, we could in principle access the entire 360$^\circ$ region around the star for discovery. \\
Long ADI sequences are still difficult to generate because of computational time. We are focusing on increasing the number of images to compare our simulations to the real amount of data taken during an on sky observation and also because we are confident that the post processing code performance will become more and more successful by using more images. \\
We also analysed some low Strehl cases, the gain of the pipeline in comparison to a simple ADI is confirmed, even if the advantage is lower and focus on the large separation from the host star.
This result and the gain of 1 magnitude in high and medium Strehl case in different coronagraph configurations encourage to implement the code on sky images to verify and adjust the pipeline for the analysis of the scientific images of the SHARK-NIR. 



\begin{thebibliography}{}

\end{thebibliography}


\begin{thebibliography}{10}

\bibitem{2014ebi..confP4.74F}
Farinato, J., Baffa, C., Carbonaro, L., Dima, M., Esposito, S., Giallongo, E.,
  Greggio, D., Hinz, P., Lisi, F., Magrin, D., Pedichini, F., Pinna, E.,
  Ragazzoni, R., and Stangalini, M., ``The nir arm of shark (system for
  coronagraphy with high order adaptive optics from r to k band),'' in [{\em
  Search for Life Beyond the Solar System. Exoplanets, Biosignatures 
  Instruments}{\nolinebreak\hspace{0.1em}]},   P4.74 (2014).

\bibitem{2015IJAsB..14..365F}
Farinato, J., Baffa, C., Baruffolo, A., Bergomi, M., Carbonaro, L., Carlotti,
  A., Centrone, M., Codona, J., Dima, M., Esposito, S., Fantinel, D., Farisato,
  G., Gaessler, W., Giallongo, E., Greggio, D., Hinz, P., Lisi, F., Magrin, D.,
  Marafatto, L., Pedichini, F., Pinna, E., Puglisi, A., Ragazzoni, R.,
  Salasnich, B., Stangalini, M., Verinaud, C., and Viotto, V., ``The nir arm of
  shark: System for coronagraphy with high-order adaptive optics from r to k
  bands,'' {\em International Journal of Astrobiology}~{\bf 14},  365--373
  (2015).

\bibitem{2016SPIE.9909E..31F}
Farinato, J., Bacciotti, F., Baffa, C., Baruffolo, A., Bergomi, M., Bongiorno,
  A., Carbonaro, L., Carolo, E., Carlotti, A., Centrone, M., Close, L.,
  De~Pascale, M., Dima, M., D'Orazi, V., Esposito, S., Fantinel, D., Farisato,
  G., Gaessler, W., Giallongo, E., Greggio, D., Guyon, O., Hinz, P., Lisi, F.,
  Magrin, D., Marafatto, L., Mohr, L., Montoya, M., Pedichini, F., Pinna, E.,
  Puglisi, A., Ragazzoni, R., Salasnich, B., Stangalini, M., Vassallo, D.,
  Verinaud, C., and Viotto, V., ``Shark-nir: from k-band to a key instrument, a
  status update,'' {\em Proc. SPIE} {\bf 9909},  990931 (2016).

\bibitem{2016SPIE.9911E..27V}
Viotto, V., Farinato, J., Greggio, D., Vassallo, D., Carolo, E., Baruffolo, A.,
  Bergomi, M., Carlotti, A., De~Pascale, M., D'Orazi, V., Fantinel, D., Magrin,
  D., Marafatto, L., Mohr, L., Ragazzoni, R., Salasnich, B., and Verinaud, C.,
  ``Shark-nir system design analysis overview,'' {\em Proc. SPIE} {\bf 9911},
  991127 (2016).

\bibitem{adi}
Ren, D., Dou, J., Zhang, X., and Zhu, Y., ``Speckle noise subtraction and
  suppression with adaptive optics coronagraphic imaging,'' {\em The
  Astrophysical Journal}~{\bf 753},  99 (2012).

\bibitem{Marois}
Marois, C., Lafreniere, D., Doyon, R., Macintosh, B., and Nadeau, D., ``Angular
  {D}ifferential {I}maging: A {P}owerful {H}igh-{C}ontrast {I}maging
  {T}echnique,'' {\em The Astrophysical Journal}~{\bf 641},  556--564 (2006).

\bibitem{pca}
Wall, M.~E., Rechtsteiner, A., and M., R.~L., ``Singular {V}alue
  {D}ecomposition and {P}rincipal {C}omponent {A}nalysis,'' {\em ArXiv Physics
  e-prints}~{\bf http://adsabs.harvard.edu/abs/2002physics...8101W} (2002).

\bibitem{pca2}
Amara, A. and Quanz, S., ``Pynpoint: an image processing package for finding
  exoplanets,'' {\em Monthly Notices of the Royal Astronomical Society}~{\bf
  427},  948--955 (2012).

\bibitem{carolo}
Carolo, E., Vassallo, D., Farinato, J., Bergomi, M., Bonavita, M., Carlotti,
  A., D'Orazi, V., Greggio, D., Magrin, D., Mesa, D., Pinna, E., Puglisi, A.,
  Stangalini, M., Verinaud, C., and Viotto, V., ``A comparison between
  different coronagraphic data reduction techniques,'' {\em Proc. SPIE} {\bf
  9909},  99097Q (2016).

\bibitem{Vassallo}
Vassallo, D., Carolo, E., Farinato, J., Bergomi, M., Bonavita, M., Carlotti,
  A., D'Orazi, V., Greggio, D., Magrin, D., Mesa, D., Pinna, E., Puglisi, A.,
  Stangalini, M., Verinaud, C., and Viotto, V., ``An extensive coronagraphic
  simulation applied to {LBT},'' {\em Proc. SPIE} {\bf 9911},  99110Y (2016).

\bibitem{Sphere}
Vigan, A., Gry, C., Salter, G., Mesa, D., Homeier, D., Moutou, C., and Allard,
  F., ``High-contrast imaging of {S}irius {A} with {VLT/SPHERE}: looking for
  giant planets down to one astronomical unit,'' {\em Monthly Notices of the
  Royal Astronomical Society}~{\bf 454},  129--143 (2015).

\bibitem{spheretec}
Beuzit, J., Boccaletti, A., Feldt, M., Dohlen, K., Mouillet, D., Puget, P.,
  Wildi, F., Abe, L., Antichi, J., Baruffolo, A., Baudoz, P., Carbillet, M.,
  Charton, J., Claudi, R., Desidera, S., Downing, M., Fabron, C., Feautrier,
  P., Fedrigo, E., Fusco, T., Gach, J., Giro, E., Gratton, R., Henning, T.,
  Hubin, N., Joos, F., Kasper, M., Lagrange, A., Langlois, M., Lenzen, R.,
  Moutou, C., Pavlov, A., Petit, C., Pragt, J., Rabou, P., Rigal, F., Rochat,
  S., Roelfsema, R., Rousset, G., Saisse, M., Schmid, H., Stadler, E.,
  Thalmann, C., Turatto, M., Udry, S., Vakili, F., Vigan, A., and Waters, R.,
  ``Direct detection of giant extrasolar planets with {SPHERE} on the {VLT},''
  {\em Astronomical Society of the Pacific workshop Proc.}~{\bf 430},  231
  (2010).

\bibitem{lyot}
Lyot, B., ``The study of the solar corona and prominences without eclipses,''
  {\em Monthly Notices of the Royal Astronomical Society}~{\bf 99},  580
  (1939).

\bibitem{sp}
Carlotti, A., Kasdin, N., Vanderbei, R., , and Delorme, J., ``Optimized shaped
  pupil masks for pupil with obscuration,'' {\em Proc. SPIE} {\bf 8442},  54
  (2012).

\bibitem{fqpm}
Rouan, D., Riaud, P., Boccaletti, A., Cl{\'e}net, Y., and Labeyrie, A., ``The
  four-quadrant phase-mask coronagraph. i. principle,'' {\em The Astronomical
  Society of the Pacific}~{\bf 112},  1479--1486 (2000).

\end{thebibliography}
\bibliographystyle{spiebib} 

\end{document}